\documentclass[12pt]{article}
\usepackage{amssymb,amsmath,epsfig}
\usepackage{graphicx}
\usepackage{cite}
\usepackage{subfig}

\numberwithin{equation}{section}
 \allowdisplaybreaks

\begin{document}
\title{\bf Existence of Compact Structures in $f(R,T)$ Gravity}

\author{Z. Yousaf$^1$ \thanks{zeeshan.math@pu.edu.pk}, M. Zaeem-ul-Haq Bhatti \thanks{mzaeem.math@pu.edu.pk} and
M. Ilyas$^2$ \thanks{ilyas\_mia@yahoo.com}\\
$^1$ Department of Mathematics, University of the Punjab,\\
Quaid-i-Azam Campus, Lahore-54590, Pakistan\\
$^2$ Centre for High Energy Physics, University of the Punjab,\\
Quaid-i-Azam Campus, Lahore-54590, Pakistan}

\date{}

\maketitle
\begin{abstract}
The present paper is devoted to investigate the possible emergence of relativistic compact stellar objects through modified $f(R,T)$ gravity. For anisotropic matter distribution, we used Krori and Barura solutions and two notable and viable $f(R,T)$ gravity formulations. By choosing particular observational data, we determine the values of constant in solutions for three relativistic compact star candidates. We have presented some physical behavior of these relativistic compact stellar objects and some aspects like energy density, radial as well as transverse pressure, their evolution, stability, Eos parameters, measure of anisotropy and energy conditions.
\end{abstract}

\section{Introduction}

General relativity (GR) is considered as the most fruitful theory
for understanding the evolution of universe and its hidden secrets,
yet the evidence of dark matter (DM) and the cosmic accelerating
nature of spacetime put some challenges on this
\cite{zs1,zs2,zi1,zi2,zi3,zi4,zi5}. The Einstein's GR explained the
cosmological phenomena in a regime of weak field, while some
modifications may be needed to study the strong fields in the
scenario of accelerating expansion of the universe. In this
direction, Qadir \emph{et al.} \cite{zs3} reinforced the requirement
of the modified relativistic dynamics and indicated that this
modification may help to settle down the problems related to DM and
quantum gravity. As a result, many techniques were used like by
introducing the cosmological constant as well as the modified
theories from time to time.

Modified gravitational theories (MGTs) are actually the
generalization of GR in which function of the Ricci scalar ($R$) is
substituted in the Einstein-Hilbert action. These modified gravity
theories are dubbed with the names, Einstein-$\Lambda$ \cite{kkk1},
$f(R)$ \cite{z1fr} ($R$ is the Ricci scalar), $f(R,T)$ \cite{z2frt}
($T$ is the trace of energy momentum tensor), $f(G)$ \cite{martin3}
($G$ is the Gauss-Bonnet term) and $f(R,T,R_{\xi\pi} T^{\xi\pi})$
gravity \cite{z3frtrmn}. In the recent times, Nojiri \emph{et al.}
\cite{bin4} presented various mathematical techniques to understand
burning issues of cosmos related to bouncing cosmos. They asserted
that gravity mediated by $f(R)$ and $f(G)$ theories could be used to
realize many hidden secrets of our universe. Once can observe the
pity good agreement results between the cosmological models in MGTs
and the observational data \cite{8,martin2,10,11}. The $f(R,T)$
gravity is one of the MGTs, in which the $f(R)$ is replaced with the
function of $R$ and $T$ \cite{12}. It is claimed that the evidence
behind the dependence of $T$ may come from the presence of imperfect
fluid or it may be some kinds of quantum effects (for further
reviews on DE and MGTs, see, for
instance,~\cite{R2,R3,R4,R5,R6,R7,R8,R9,R10,k3,k4,k5,k6}).

In $f(R,T)$ gravity, many cosmological applications were discussed
in \cite{13,14,16,20,23,24}. From literature, some of them are, The
non-static line element for collapsing of spherical body having
anisotropic fluid were discussed in \cite{27}. The static spherical
wormhole solutions were found in \cite{28,29}. Furthermore, the
perturbation techniques were used in study of spherical stars
\cite{astr1}. The effects on gravitational lensing due to $f(R,T)$
gravity were discussed in \cite{33}. The spherical equilibrium theme
of polytropic and strange stars were investigated in \cite{34}.
Houndjo \cite{ya10} constructed few observationally notable cosmic
models in $f(R,T)$ gravity for studying matter dominated era of the
expanding universe. Baffou \textit{et al.} \cite{ya13} applied
perturbation on the spacetimes of de-Sitter and power law models in
order to explore some cosmic viability bounds.

Bamba \emph{et al.} \cite{z5d} analyzed the effects of higher
degrees of freedom coming from MGT on the dynamical features of our
accelerating cosmos. Bamba \emph{et al.} \cite{grg1} further checked
the viability regimes on the parameters of $f(G)$ gravity models and
presented some mathematically consistent cosmic zones. The stability
of gravitational evolving stellar bodies have been investigated in
few models of $f(R)$ gravity by \cite{mart5,mart7}. Das \emph{et
al.} \cite{das1} calculated exact relativistic models of spherical
interiors in MGT and discussed the physical implications of their
results on compact stars.

Yousaf and his collaborators examined the role of various curvature
invariant functions on the existence as well as stability of the
planar \cite{7p}, spherical \cite{7s} and cylindrical \cite{7c}
geometries. Sahoo with his coworkers \cite{zs19} studied the
viability of the spatially regular cosmos along with some other
cosmological aspects in $f(R,T)$ gravity. Moraes \emph{et al.}
\cite{9} worked out the stability of some well-known compact stars
by computing their corresponding hydrostatic equations in
$f(R,T)=R+2\lambda T$ gravity.

The exploration on the existence of self-gravitating compact stars
have always been a source of great attention among gravitational
physicists. \cite{35,36,37,38,44,45,46,46a}. In this direction, many
researchers reconstructed different models for the study of
anisotropic relativistic compact stars. Various physical properties,
like masses of compact stars, radii, stability etc. and the moment
of inertia of neutron stars were studied with the comparison
developed with GR and other MGTs \cite{47}. Virbhadra \emph{et al.}
\cite{vir1,vir2} discussed the relation between naked singularity
and black holes formation accompanied with a well-consistent
mathematical stand point. Egeland \cite{48} performed the modeling
of neutron star by examining some mass radius relations and
concluded that $\Lambda$ should exist to justify the vacuum density.
Sharif and Yousaf \cite{gerg1} also found these relations and
checked the existence of compact structures in the platform of MGTs.
Bhatti \emph{et al.} \cite{bhatti1} calculated dark dynamical
variables and checked the dynamics of compact stars with the help of
these variables. Recently, Yousaf \emph{et al.} \cite{epjc1}
investigated the stability of three different compact structures in
the presence of dark sources terms mediated by quadratic, cubic and
exponential $f(R)$ formulations.

Here, we use the modified $f(R,T)$ gravity, which can be considered
as more well established theory than that of cosmological constant.
To explore the formation of relativistic anisotropic compact stellar
objects in $f(R,T)$ gravity, we use two such a viable $f(R,T)$
models for three candidates of strange stars, i,e., Her X-1,
SAXJ1808.4-3658 and 4U1820-30. We shall label these stars with CS1,
CS2 and CS3, respectively. We find the solutions of relativistic
anisotropic stellar bodies in $f(R,T)$ gravity, that enforce some
constraints on the cosmic model parameters. In very next section, we
formulate $f(R,T)$ equation of motions, after having the background
of $f(R,T)$ gravity, we take anisotropic matter contents for static
spherical star geometry. In section \textbf{3}, we demonstrate some
of the physical viable models in $f(R,T)$ gravity. In section
\textbf{4}, we explore the solutions and discussed the physical
properties of relativistic compact stars through graphical
illustrations. At the end, we finalize our results.

\section{$f(R,T)$ gravity}

The action for $f(R,T)$ gravity can be written as
\begin{equation}\label{action1}
I=\int{d{x^4}\sqrt{-g}\left[{f(R,T)+{L_{matter}}}\right]},
\end{equation}
where $g$ and $L_{matter}$ indicate determinant of metric tensor and
the matter Lagrangian, respectively. By varying the above equation
with metric tensor, we get
\begin{align}\nonumber
&{R_{\xi \pi}}{f_R}-\frac{1}{2}{g_{\xi \pi}}f + \left(
{{g_{\xi \pi}}{\nabla ^\xi }{\nabla _\mu } - {\nabla _\mu
}{\nabla _\pi }} \right){f_R}\\\label{fequation1} &= {T_{\xi \pi}}
-{f_T}{\Theta _{\xi \pi }}-{f_T}{T_{\xi \pi}},
\end{align}
where the notations ${f_R},~{\nabla_\gamma}$ and ${f_T}$ are operators for covariant and
partial derivatives of their arguments, respectively, while ${\Theta _{\mu \nu }}$ can be expressed through the stress-energy tensor
($T_{\gamma\delta }$) as
\begin{equation}\label{2}
{\Theta _{\xi \pi }} = \frac{{{g^{\alpha \beta }}\delta {T_{\alpha
\beta }}}}{{\delta {g^{\xi \pi }}}} = - 2{T_{\xi \pi }} + {g_{\xi \pi }}{L_m} - 2{g^{\alpha \beta }}\frac{{{\partial
^2}{L_m}}}{{\partial {g^{\xi \pi }}\partial {g^{\alpha \beta }}}}.
\end{equation}
The aim of this work is to check the different properties of some of the compact stars with locally anisotropic pressure. Now, we consider
that the spherical distribution of geometric system
is coupled with the following relativistic fluid
\begin{align}\label{2l}
{T_{\mu \nu }} = (\rho  + {P_t}){V_\mu
}{V_\nu } - {P_t}{g_{\mu \nu }} + \Pi{X_\mu }{X_\nu },
\end{align}
in which $\Pi\equiv P_r-P_\bot$ and $\rho$ is the matter energy density and ${V_\gamma}$ and $X_\gamma$
are the 4-vectors corresponding to fluid and radial directions, respectively. Under the non-tilted
coordinate system, the four vectors satisfy ${V^\alpha }{V_\alpha }=1$ and ${X^\beta }{X_\beta }=-1$ relations.
Equation (\ref{2}) can be written, after choosing $L_m=\rho$, as
$${\Theta _{\alpha \beta }} =  - 2{T_{\alpha \beta }} + \rho {g_{\alpha \beta }}$$
Then, the field equations (\ref{fequation1}) boil down to
\begin{equation}
{R_{\xi \pi }} - \frac{1}{2}R{g_{\xi \pi }} = T_{\xi \pi
}^{\textrm{eff}},
\end{equation}
where $T_{\xi \pi
}^{\textrm{eff}}$ is dubbed with the effective stress energy tensor for $f(R,T)$ gravity whose mathematical formulations is given by
\begin{equation}
\begin{gathered}
T_{\xi \pi }^{\textrm{eff}} = \frac{1}{{{f_R}}}\left[ {(1 + {f_T})}
\right.{T_{\xi \pi }} - \rho {g_{\xi \pi }}{f_R} + \frac{1}{2}(f
- R{f_R}){g_{\xi \pi }} + ({\nabla _\xi }{\nabla _\pi } -
{g_{\xi \pi }}{\nabla ^\alpha }{\nabla _\alpha })\left. {f_R}
\right].
\end{gathered}
\end{equation}

Now, we wish to consider the diagonally symmetric static form of spherically symmetric spacetime as
\begin{equation}\label{zz7}
d{s^2} = {e^{a}}d{t^2} - {e^{b}}d{r^2} - {r^2}\left( {d{\theta ^2} + {{\sin }^2}\theta d{\phi ^2}} \right),
\end{equation}
where metric coefficients $a$ and $b$ are the $r$ dependent
functions. Our aim is to analyze the role of anisotropicity in the
modeling of some stellar toy models. So, we assume that
Eq.(\ref{zz7}) is composed of the locally anisotropic fluid content
given in Eq.(\ref{2l}). The $f(R,T)$ equations of motion
(\ref{fequation1}) for the geometry (\ref{zz7}) and fluid (\ref{2l})
yield
\begin{align}\label{ro}
\rho&= {e^{ - b}}\left[ {\left\{ \frac{1}{4}a'{{(r)}^2}+{\frac{{a''}}{2}-
\frac{b'}{4}a'+ \frac{{a'}}{r}}
\right\}{f_R}\left( {R,T} \right)}+ \left( {\frac{{b'(r)}}{2} -
\frac{2}{r}} \right) \right.\left.{
{f_R}' - \frac{{f_R}}{2}{e^{b}}-{f_R}{{^\prime }^\prime }}
\right],\\\label{pr}  P_r &= \frac{{{e^{ - b}}}}{{\left( {1 +
{f_T}\left( {R,T} \right)} \right)}}\left[ {\left\{ { -
\frac{{a''}}{2} - \frac{b'}{4}a' +
\frac{{b'}}{r}}-\frac{1}{4}a'^2 \right\}{f_R}}
\right.\left. {+ \left( {\frac{{a'}}{2} +
\frac{2}{r}} \right){f_R}^\prime +
\frac{{e^{b}}}{2}{f_R}} \right] -
\frac{{\rho {f_T}}}{{\left( {1 + {f_T}} \right)}},
\\\nonumber P_t&= \frac{{{e^{ - b}}}}{{\left( {1 + {f_T}} \right)}}\left[ {\left\{ {e^{b}}+{\frac{r}{2}\left( {b'-
a'} \right) - 1} \right\}\frac{{{f_R}}}{{{r^2}}}}+ \left\{-\frac{1}{2}( {b'-a'})\right.
\right. \left.\left.+ \frac{1}{r} \right\} {
{f_R}^\prime +{f_R}^{\prime \prime }+ \frac{{f_R}}{2}{e^{b}}
}
\right]\\\label{pt}&-\frac{{\rho {f_T}}}{{\left( {1 + {f_T}} \right)}},
\end{align}
where the over prime indicates $\partial/\partial r$ operator.

In order to analyse the impact of $f(R,T)$ gravity on the construction of stellar models, we
assume the separable form of $R$ and $T$
in $f(R,T)$ model as follows
\begin{align}\label{mm1}
f(R,T)=f_i(R)+g(T).
\end{align}
This choice of separable $R$ and $T$ can be regarded as a possible linear corrections in the well-known $f(R)$
theory. By choosing $f_i(R)$ from \cite{martin2} along with the linear combination of $g(T)$, 
the viable $f(R,T)$ model can be designed. Therefore, we suppose $g(T)=\epsilon T$ in which $\epsilon$
is a very small positive number. In this context, Eqs.(\ref{ro})-(\ref{pt})
provide
\begin{align}\label{z1}
\rho&= \frac{1}{{2(1 + 2\epsilon )}}\left[ {\frac{{(2 + 5\epsilon
)}}{{(1 +\epsilon )}}{\varphi_1} + \epsilon {\varphi_2} + 2\epsilon {\varphi_3}}
\right],\\\label{z2} P_r&= \frac{{ - 1}}{{2(2\epsilon  + 1)}}\left[
{\frac{\epsilon }{{\left( {\epsilon  +1} \right)}}{\varphi_1} - (2 +
3\epsilon ){\varphi_2} + 2\epsilon {\varphi_3}} \right],\\\label{z3} P_t&=
\frac{{ - 1}}{{2(1 + 2\epsilon )}}\left[ {\frac{\epsilon }{{(1
+\epsilon )}}{\varphi_1} + \epsilon {\varphi_2} - 2(1+ \epsilon  ){\varphi_3}}
\right],
\end{align}
where
\begin{align}\nonumber
{\varphi_1} &= {e^{ - b}}\left[ {\left( {\frac{a'^2}{4}+\frac{{a''}}{2} -
\frac{b'}{4}a' + \frac{{a'}}{r}}
\right){f_R}}+ \left( {\frac{{b'}}{2} - \frac{2}{r}} \right)
\right. \left. { \times{f_R}^\prime
-\frac{{f_R}}{2}{e^{b}} - {f_R}^{\prime \prime }} \right],
\hfill
\\\nonumber
{\varphi_2}&= \frac{{{e^{ - b}}}}{{\left( {1 + \epsilon } \right)}}
\left[ {\left( {\frac{{b'}}{r} - \frac{{a''}}{2} - \frac{b'}{4}a' -
\frac{1}{4}a'^2} \right){f_R}}+ \left(
{\frac{2}{r}+\frac{{a'}}{2}} \right)\right. \left. { \times{f_R}^\prime + \frac{1}{2}{e^{b}}{f_R}} \right],
\hfill
\\\nonumber {\varphi_3}&= \frac{{{e^{ - b}}}}{{\left( {1 + \epsilon }
\right)}} \left[ {\left( {{e^{b}}-\frac{1}{2}r\left( {a'-b'} \right) - 1} \right)\frac{{{f_R}}}{{{r^2}}}}+ \left\{
{\frac{1}{r}-\frac{1}{2}\left( {b'-a'} \right)}
\right\} {{f_R}^\prime  + \frac{{f_R}}{2}{e^{b}} +
{f_R}^{\prime \prime }} \right].
\end{align}
We write $a$ and $b$ as a combinations of radial coordinates
suggested by Krori and Barua \cite{zs43}. They proposed the specific
forms of these functions as $a(r) =B r^2 +C$ and $b(r)=A r^2$, where
$A,~B$ and $C$ are the three constant numbers. One can find the
values of this triplet $(A,~B,~C)$ by considering some appropriate
boundary conditions. Then, Eqs.(\ref{z1})-(\ref{z3}) yield
\begin{align}\nonumber
\rho &= \frac{{{{\rm{e}}^{ - A{r^2}}}}}{{2{r^2}\left( {1 + \epsilon } \right)\left( {1 + 2\epsilon }
\right)}}[ - {{\rm{e}}^{A{r^2}}}{r^2}\left( {1 + \epsilon } \right)f_i + (2( - 1 - 2\epsilon  + 3B{r^2}
\epsilon + {B^2}{r^4}\epsilon + {{\rm{e}}^{A{r^2}}} \\\nonumber
 &\times\left( {1 + 2\epsilon } \right) + A{r^2}\left( {2 + 4\epsilon  - B{r^2}\epsilon } \right))
 + {{\rm{e}}^{A{r^2}}}{r^2}(1 + \epsilon )R)f_i'+ r(( - 4 - 6\epsilon  + 3B{r^2}\epsilon \\\label{ro1}
 & + A{r^2}(2 + 3\epsilon )){R^\prime }f_i''-r\left( {2 + 3\epsilon } \right)
 {R^{\prime \prime }}f_i''- r\left( {2 + 3\epsilon } \right){R^\prime }^2f_i''')],\\\nonumber
{P_r}&= \frac{{{{\rm{e}}^{ - A{r^2}}}}}{{2{r^2}\left( {1 + \epsilon } \right)\left( {1 + 2\epsilon }
\right)}}[{{\rm{e}}^{A{r^2}}}{r^2}\left( {1 + \epsilon } \right)f_i - (2( - 1 - 2\epsilon  + {B^2}{r^4}
\epsilon + {{\rm{e}}^{A{r^2}}}(1 + 2\epsilon ) \\\nonumber
 &- B{r^2}\left( {2 + \epsilon  + A{r^2}\epsilon } \right)) + {{\rm{e}}^{A{r^2}}}{r^2}\left( {1 + \epsilon } \right)R)f_i'
 + r[\{4 + 6\epsilon  + A{r^2}\epsilon  + B{r^2}( 2 + \epsilon ) \}\\\label{pr1}
& \times R'f_i'' - r\epsilon R''f_i'' - r\epsilon R'^2f_i''']],\\\nonumber
{P_t} &= \frac{{{{\rm{e}}^{ - A{r^2}}}}}{{2{r^2}\left( {1 + \epsilon } \right)\left( {1 + 2\epsilon }
\right)}}[{{\rm{e}}^{A{r^2}}}r\left( {1 + \epsilon } \right)fi - r[2\{ - B(2 + \epsilon + B{r^2}(1 +
\epsilon )) + A(1 + 2\epsilon\\\nonumber
& + B{r^2}(1 + \epsilon ))\}+ {{\rm{e}}^{A{r^2}}}\left( {1 + \epsilon } \right)R]f_i'+2{R^\prime }
f{i^{\prime \prime }}- 2A{r^2}{R^\prime }f{i^{\prime \prime }}+ 2B{r^2}{R^\prime }f{i^{\prime \prime }}
+2\epsilon R'f_i''\\\label{pt1}
&-3A{r^2}\epsilon R'f_i'' + B{r^2}\epsilon R' f_i'' + 2f_i''rR''
 + 3r\epsilon R''f_i'' + 2r{R^\prime }^2f_i'''+ 3r\epsilon R'^2f_i'''].
\end{align}

\section{Boundary Conditions}

We consider a timelike hypersurface denoted by $\Omega$ that has differentiated
interior manifold given in Eq.(\ref{zz7}) and outer geometry. The exterior region is through the
vacuum Schwarzschild spacetime as
\begin{equation}\label{ex1}
d{s^2} = \left( {1 - \frac{{2M}}{r}} \right)d{t^2}-{\left( {1 - \frac{{2M}}{r}}
\right)^{ - 1}}d{r^2} - {r^2}\left( {d{\theta ^2} + {\mathop{\rm \sin}\nolimits} {\theta ^2}d{\varphi ^2}} \right),
\end{equation}
where $M$ is the gravitating matter content. At the boundary surface $\Omega$, 
the continuity of the metric variables $g_{tt},~g_{rr}$ and
the $\partial g_{tt}/\partial r$ provide the following
constraints
\begin{align}
A &= \frac{{ - 1}}{{{R^2}}}\ln \left( {1 - \frac{{2M}}{R}} \right), \quad
B = \frac{M}{{{R^3}}}{\left( {1 - \frac{{2M}}{R}} \right)^{ - 1}},\\
C &= \ln \left( {1 - \frac{{2M}}{R}} \right) - \frac{M}{R}{\left( {1 - \frac{{2M}}{R}} \right)^{ - 1}}.
\end{align}
For the smooth and continuous matching of the manifolds, the above
constraints at $\Omega$ should be fulfilled. We now pick some
observational values of $A,~B$ and $C$ from the literature to check
the construction as well as the stability of the compact stellar
bodies. We consider, three stellar toy models Her X-1,
SAXJ1808.4-3658 and 4U1820-30 labeled with CS1, CS2 and CS3 here.
The masses of these star candidates are  $0.88$, $1.435$ and $2.25$
solar masses, respectively. All of these stellar structures satisfy
Buchdahl Bondi bound as their $2M/R$ ratios are less than $8/9$. We
further suppose that the junction conditions at the stellar core are
\cite{i1}
\begin{align}
\rho (r) = {\rho _c}\\
\frac{{da}}{{dr}}(0) = 0,\\
b(0) = 0,\\
\end{align}
where $\rho _c$ is the critical mass density.
\begin{table}[h!]
\centering
\begin{tabular}{|c| c| c| c| c| c|}
 \hline
Compact Stars (CS)  &$R(km$ & $M$ & $\mu=\frac{M}{R}$ &$B(km^{-2})$ &$A(km^{-2})$ \\ [0.5ex]
\hline\hline\
CS1  & $7.7$ & $0.88M_{\odot}$ & $0.168$ & $0.004267364618 $ & $0.006906276428 $\\ [1ex]
\hline\
CS2 & $7.07$ & $1.435M_{\odot}$ & $0.299$ & $0.01488011569$ &$0.01823156974$\\ [1ex]
\hline
CS3 & $10$ & $2.25M_{\odot}$ & $0.332$ & $0.009880952381$ & $0.01090644119 $\\ [1ex]
\hline
\end{tabular}
\caption{The observational values of the radii $R$, masses $M$, compactness $\mu$, and the constants $B$ and $A$ for CS1, CS2 and CS3.}
\label{table:1}
\end{table}

\section{Various $f(R,T)$ Models}

The aim of this section is to study some of the notable $f(R,T)$
models. We want to analyze the influences of these $f(R,T)$ models
in the construction and stability of the compact stellar candidates
in the background of some observations. We shall discuss some
realistic features of spherical systems like, compactness,
stability, evolution of energy density with radial coordinate, the
measurement of anisotropic pressure evolution and the different
energy conditions. The results obtained such investigations may
provide some hidden realities corresponding to the both theoretical
and astrophysical regimes. Equation (\ref{mm1}) provides
\begin{equation}\label{modelfRT}
f(R,T)=fi(R)+ \epsilon T,
\end{equation}
where $i=1,~2$. There has been very interesting results which reveal
that the inclusion of extra dark energy/DM components mediated from
alternative theories could bring some exciting results. For
instance, collapse time, existence of more compact structure,
stability, phenomenon of core formation and above all can be well
influenced by these dark source terms unlike GR. The reexamination
of GR problems in alternatives theories may be helpful to shed some
light on the models viability and their usefulness on physical
grounds. The cross-examinations may present both quantitative and
qualitative different consequences than that of GR.

It is of particular interest for many researchers to explore the
predictions as well as the outcomes of modified gravitational
theories like $f(R,T)$ theory concerning the existence of stellar
structure and their stability. The $f(R,T)$ gravity can be treated
as a mathematical tool to examine various unknown features of
gravitational dynamics at large scales. Schaefer and Koyama
\cite{rev1} generalized their study of gravitational collapse of
spherical structures in the realm of modified gravity and
Birkhoff-theorem and found enhanced cluster merging rates as well as
overdensed populations of relativistic structures (due to modified
gravity). Capozziello et al. \cite{rev2} computed modified
Lan\'{e}-Emden expression with metric $f(R)$ corrections and
Newtonian approximation. They get some exceptional results of
density and pressure distribution in the analysis of the hydrostatic
scenario of few celestial bodies. Cembranos et al. \cite{rev3}
studied stellar structure formation in the background of modified
gravity and found comparatively higher contraction in the collapsing
rate of spherical systems at its initial stages unlike GR.

Astashenok et al.\cite{rev4} examined the impact of few modified
gravity formulations on the existence of compact structure and
inferred that there exists some models whose corrections could lead
arena of having relatively more compact stars than in GR. Yousaf and
Bhatti \cite{mnras1} studied the influences of modified dark source
terms on mass-radius relationships for compact stars and concluded
that some extended gravities mediated by Lagrangians $\alpha R^2$
and $\frac{\alpha R^2-\beta R}{1+\gamma R}$ are likely to host
supermassive relativistic systems with comparatively smaller radii
than in GR. Resco et al. \cite{rev5} numerically calculated apparent
masses of neutron star models in modified gravity and inferred that
generically some modified extra degrees of freedom likely to keep
significantly massive neutron star with smaller radii than in GR.
Such type of investigations could provide theoretical
well-consistent way to handle and study classes of massive and super
massive structures at large scales. Bamba et al. \cite{rev6} claimed
that modified gravitational dynamics could provide an additional
platform in few regions of spacetime that could lead to stable
configurations of relativistic system. In this paper, we found
relatively stellar bodies with much higher densities than that found
in GR  \cite{rev7}  as seen from Fig.(\ref{roc}).

In the following we discuss two different $f(R,T)$ models.

\subsection{Model 1}

Firstly, we take the tanh modification of the Ricci scalar function that was proposed by \cite{tan1}. In this respect, the $f(R,T)$ gravity model (\ref{modelfRT}) give
\begin{equation}\label{model1}
f(R,T)= R-\alpha \hat{R} \tanh \left(\frac{R}{\hat{R}}\right)+\epsilon T,
\end{equation}
where $\alpha\in \mathbb{R}^+$ and $\hat{R}\in \mathbb{R}^+$ in which $\mathbb{R}^+$ denotes the set of positive real numbers. On setting $\alpha= 0$, the dynamics of Einstein's gravity can be discussed.

\subsection{Model 2}

Next, we consider another viable formulations of $f(R)$ gravity that was first suggested by \cite{suji1}. Under this context, Eq.(\ref{modelfRT}) becomes
\begin{equation}\label{model3}
f(R,T)= R+\gamma \hat{R} \left((1+\frac{R^2}{\hat{R^2}})^{-q}-1\right)+\epsilon T,
\end{equation}
where $\hat{R},~\gamma$ and $q$ are the free non zero and non negative parameters.

In the coming section, we will use these two viable model for further investigation of compact stars.

\section{Some Physical Properties}

This section is devoted to analyze various aspects of some compact
stellar toy models. We assume three different star distributions,
i.e., CS1, CS2, CS3 as well as two observationally consistent
$f(R,T)$ models. Upon substituting this data in
Eqs.(\ref{ro})-(\ref{pt}), we get values of matter variables in the
form of four parameters, $\alpha,~\hat{R},~q$ and $\epsilon$. Then,
we study some physical features to obtain the realistic
configurations of compact stellar structures (shown in Table
\ref{table:1}). The comparison between our theoretical outcomes and
the observational data may provide the strong evidences for $f(R,T)$
models. We shall show our results with the help of plots.

\subsection{Density and Pressure Evolutions}

\begin{figure} \centering
\epsfig{file=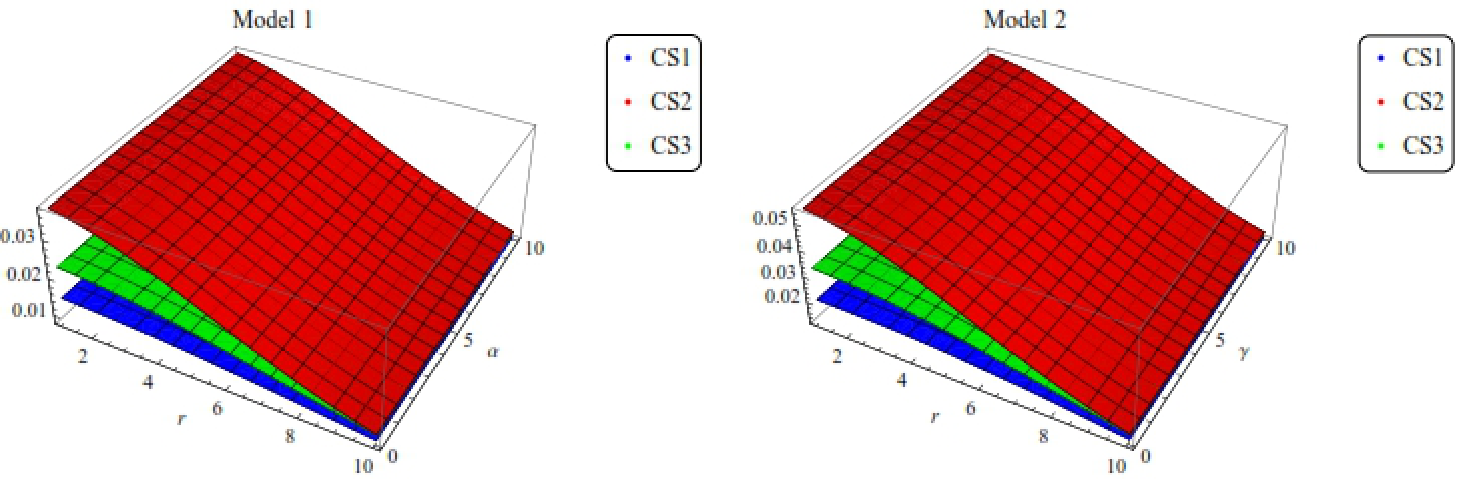,width=0.8\linewidth} \caption{Graph for the
density ($km^{-2}$) evolution with both $f(R,T)$ models.}\label{roc}
\epsfig{file=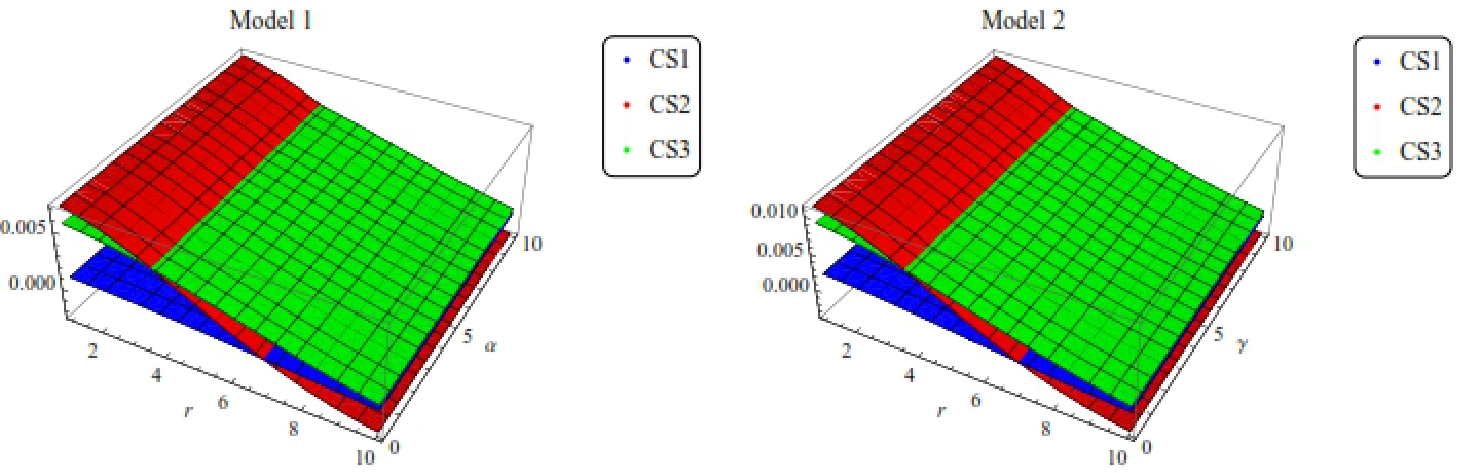,width=0.8\linewidth} \caption{Graph for the
radial pressure ($km^{-2}$) evolution with both $f(R)$
models.}\label{prc}
\end{figure}

\begin{figure} \centering
\epsfig{file=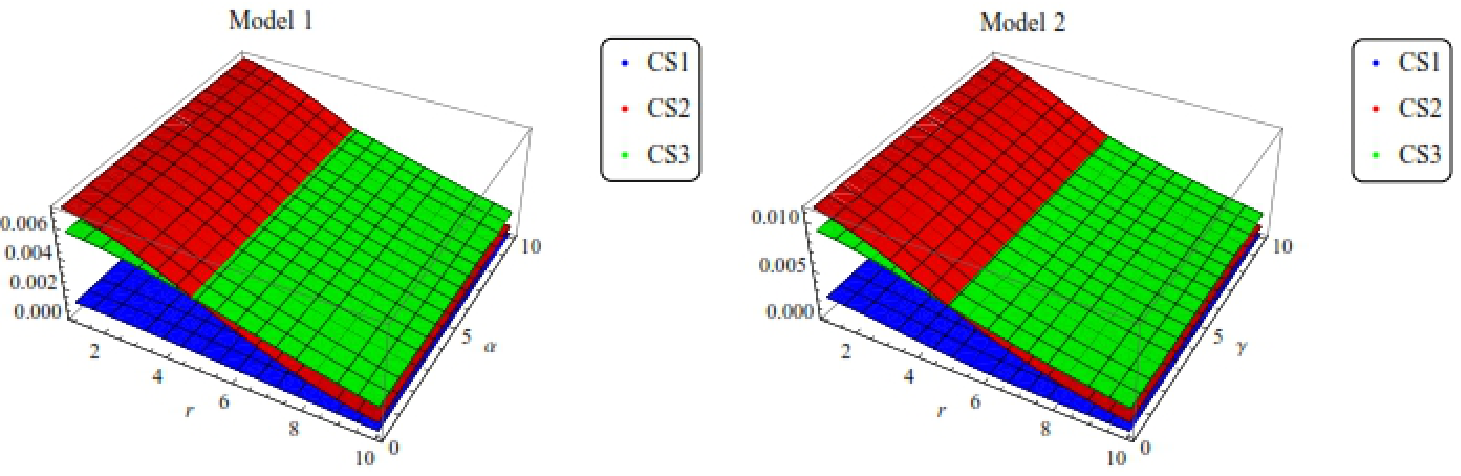,width=0.8\linewidth} \caption{Graph for the
transverse pressure ($km^{-2}$) evolution with both $f(R,T)$
models.}\label{ptc}
\end{figure}

\begin{figure} \centering
\epsfig{file=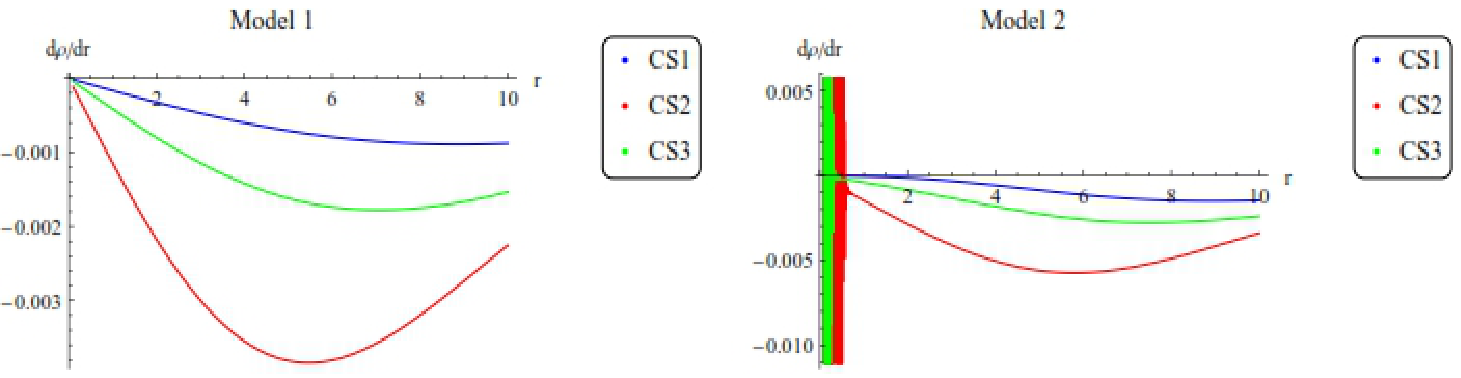,width=0.6\linewidth}
\caption{Graph describing $d\rho/dr$ with respect to $r$ for both $f(R)$ models.}\label{dro}
\epsfig{file=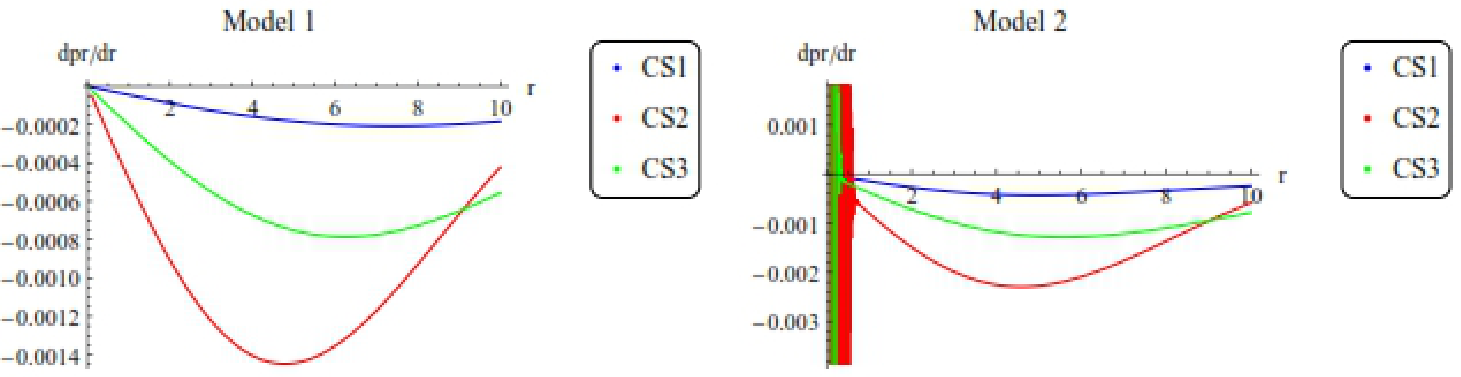,width=0.6\linewidth}
\caption{Graph describing $dp_r/dr$ with increasing $r$ for both $f(R)$ models.}\label{dpr}
\epsfig{file=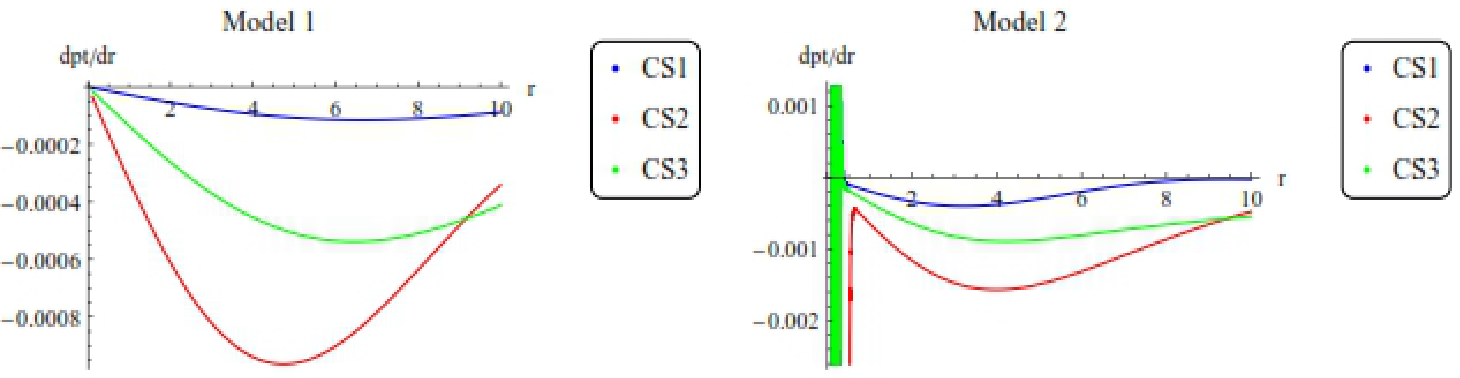,width=0.6\linewidth}
\caption{Graph describing $dp_t/dr$ with increasing $r$ for both $f(R)$ models.}\label{dpt}
\end{figure}

This section is devoted to analyze the matter parameters of all the
three stellar bodies whose variations are depending on the radial
coordinate. We check the radial variations in the anisotropic
pressure, energy density and their corresponding radial derivatives.
Now with the help of Eqs.(\ref{ro})-(\ref{pt}), we plot diagrams
(\ref{roc}) for all the three stellar models in the background of
two above mentioned $f(R,T)$ models> We infer that the behavior of
energy densities keep on increasing till the constraint
$r\rightarrow0$. This indicates the distribution of $\rho$ to be
increasing with respect to the decreasing choices of $r$. One can
say from these that our stellar toy models are of having highly
compact cores in the degrees of freedom coming from
Eqs.(\ref{model1}) and (\ref{model3}). The Figures (\ref{prc}) and
(\ref{ptc}) describe the plots
of the variations in the components of the star pressure.\\
Furthermore, Figures (\ref{dro}), (\ref{dpr}) and (\ref{dpt}) are describing the variations in
the $r$ derivative of $\rho$ and $p_i$. One can observe from these that $\frac{d\rho}{{dr}}<0,~\frac{dp_r}{{dr}}<0$
and $\frac{dp_t}{{dr}}<0$ for the given two viable models and the three toy models. Under the constraint $r=0$,
one can notice that the $r$ variations of all the matter variables vanishes, such that
$$\frac{d\rho}{{dr}}=0,$$
$$\frac{dp_r}{{dr}}=0.$$
Furthermore, the twice variations of these structural quantities have been found to be negative. These consequences
are providing a seed for the high compact profiles of such star structures around their subsequent cores, thus
indicating that the configurations of compact and dense stars do exists in the arena of $f(R,T)$ gravity.

\subsection{Energy Conditions}

\begin{figure}
\centering
\epsfig{file=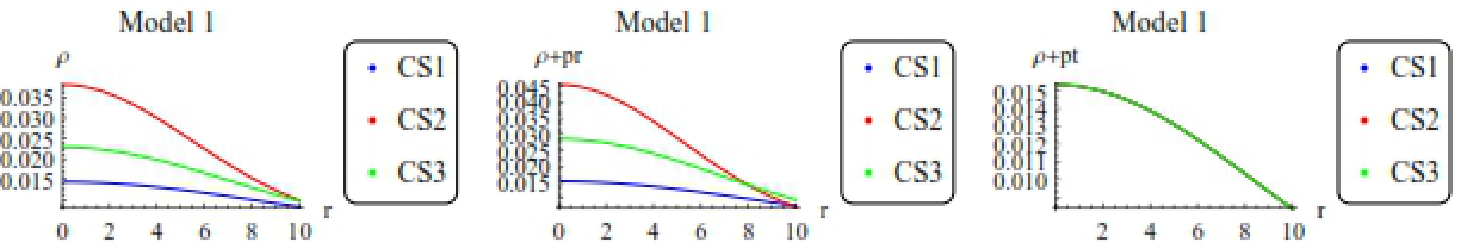,width=1\linewidth}
\epsfig{file=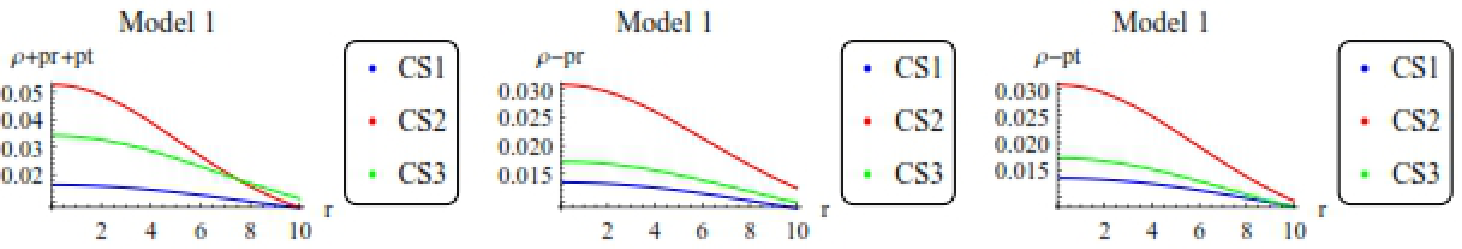,width=1\linewidth}
\caption{ECs Viability for three stellar bodies with $f(R,T)= R-\alpha \hat{R} \tanh \left(\frac{R}{\hat{R}}\right)+\epsilon T$}\label{energy1}
\end{figure}

\begin{figure}
\centering
\epsfig{file=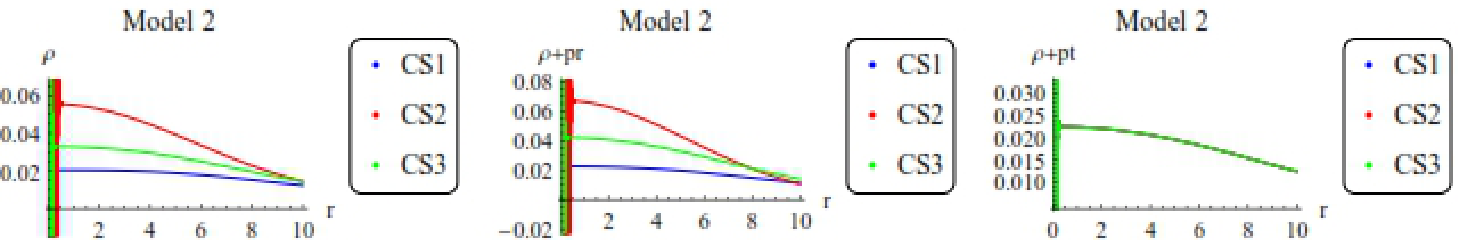,width=1\linewidth}
\epsfig{file=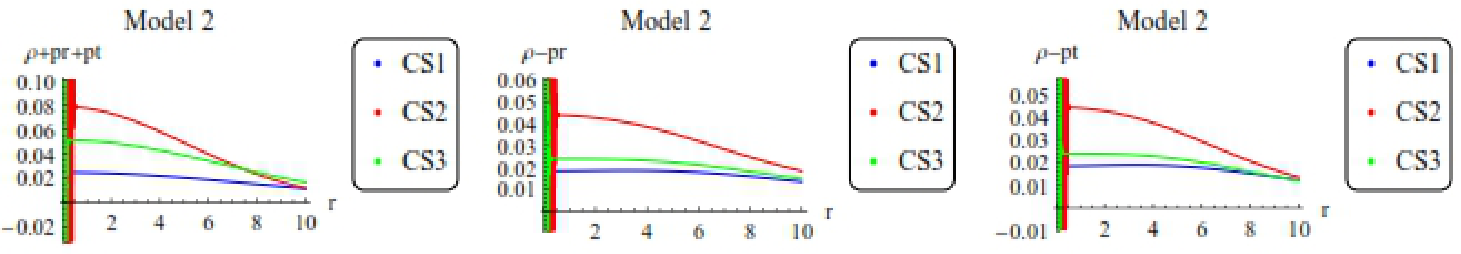,width=1\linewidth}
\caption{ECs Viability for three stellar bodies with $f(R,T)= R+\gamma \hat{R} \left((1+\frac{R^2}{\hat{R^2}})^{-q}-1\right)+\epsilon T$}\label{energy2}
\end{figure}

For a realistic matter content distribution, one need to
check some particular conditions, such a mathematical
conditions are called energy conditions (ECs). These mathematical
conditions are the coordinate-invariant. In the regime of modified
$f(R,T)$ gravity like the effective forms of density and the
anisotropic pressure, the null energy Condition (NEC)
and the weak energy conditions (WEC) are devised, respectively,
as
\begin{align}\nonumber
&\bullet\textrm{NEC}\Leftrightarrow {\rho^{\textrm{eff}}} +
{P_i^{\textrm{eff}}} \ge 0,\\\nonumber &\bullet\textrm{WEC}\Leftrightarrow
{\rho^{\textrm{eff}}} \ge 0 \text{ and } {\rho^{\textrm{eff}}} +
{P_i^{\textrm{eff}}} \ge 0,
\end{align}
while the rest of ECs, i.e., strong energy condition (SEC) and dominant energy condition (DEC) give
\begin{align}\nonumber
&\bullet\textrm{SEC}\Leftrightarrow {\rho^{\textrm{eff}}+2{P_t^{\textrm{eff}}}+
3{P_r^{\textrm{eff}}}} \ge 0 \text{ and } {\rho^{\textrm{eff}}} +
{P_i^{\textrm{eff}}} \ge 0,\\\nonumber &\bullet\textrm{DEC} \Leftrightarrow
{\rho^{\textrm{eff}}} \ge 0 \text{ and } {\rho^{\textrm{eff}}} \pm
{P_i^{\textrm{eff}}} \ge 0.
\end{align}
It has been easily be figure out from the plots shown in Figures (\ref{energy1}) and (\ref{energy2}) that
our under considerations of both models with CS1, CS2 and CS3 obey all ECs. This suggests that
anisotropic fluid content (\ref{2l}) describes the realistic source of
gravitational effects.

\subsection{TOV Equation}

\begin{figure} \centering
\epsfig{file=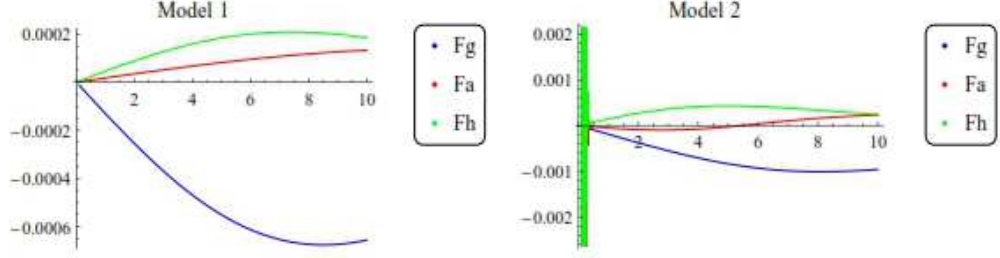,width=0.8\linewidth}
\caption{Graph describing $F_g,~F_h$ and $F_a$ with $r$ (km).}\label{eqb}
\end{figure}

A mathematical expression that provide some restrictions on the geometrical
distributions of the relativistic spherical system,
which is normally hydrostatic equilibrium and is often called
as TOV equation. For the spherical line element, this equation is written as
\begin{equation}
\frac{{d{P_r}}}{{dr}} + \frac{{a'(\rho  + {P_r})}}{2} + \frac{{2({P_r} - {P_t})}}{r} = 0,
\end{equation}
The above equation contains three
well-known interactions, i.e., the gravitational,
$(F_g)$, hydrostatic, $(F_h)$ and anisotropic, $(F_a)$ forces. Therefore, one may write above equation
as
\begin{equation}
F_g + F_h + F_a= 0.
\end{equation}
The values of these forces by using of Eqs.(\ref{ro})-(\ref{pt}) giives
\begin{align}\nonumber
F_g&\equiv-B r (\rho+P_r)=\frac{-1}{1+\epsilon}\left[2B\exp(Ar^2)r
\{2(B+A)f_i'+(B+A)R'rf_i''-R''f_i''-R'^2fi'''\}\right],\\\nonumber
F_h&\equiv-\frac{{d{p_r}}}{{dr}}=\frac{\exp(-Ar^2)}{2r^3(1+\epsilon)(1+2\epsilon)}[4\{1+2\epsilon+B^2r^4\epsilon+A^2Br^6\epsilon-\exp(Ar^2)
(1+2\epsilon)+Ar^2\\\nonumber
&\times(1+2Br^2+2\epsilon-B^2r^4\epsilon)\}fi'+r\{-r(4+3(2+Ar^2)\epsilon+Br^2(2+\epsilon))f_i'''R'^2+R'\{(2\\\nonumber
&+8Ar^2-6Br^2+4ABr^4+2\epsilon+11Ar^2\epsilon-3Br^2\epsilon+2A^2r^4\epsilon+2B^2r^4\epsilon+\exp(Ar^2)(2\\\nonumber
&+4\epsilon)
+\exp(Ar^2)r^2(1+\epsilon)R)f_i''+3r^2\epsilon R''f_i'''\}-f_i''r((4+6\epsilon+3Ar^2\epsilon+Br^2(2+\epsilon))R''\\\nonumber
&-r\epsilon R''') +r^2\epsilon R'^3fi''''\}],
\\\nonumber F_a&\equiv \frac{{2({P_r} - {P_t})}}{r}=\frac{2\exp(-Ar^2)}{r^3(1+\epsilon)}[\{\exp(Ar^2)-(1+Ar^2-Br^2)(1+Br^2)\}f_i'
+r\{-(1+Ar^2)\\\nonumber
&\times f_i'' R'+f_i''rR''+R'^2rf_i'''\}].
\end{align}
Using above relations together with observational values of $A,~B$ and $C$ from Table \ref{table:1}, we have drawn some diagrams mentioned in Figure (\ref{eqb}).
In Figure (\ref{eqb}), the left
plot is for $f(R,T)$ model (\ref{model1}) and right plot is for model (\ref{model3}).
The role of these three types of interactions with respect
to $r$ (km) has been seen in this diagram in the modeling of
compact bodies CS1, CS2 and CS3.

\subsection{Stability Analysis}

This section illustrates the stability of the compact relativistic structures.
It would be interesting to mention that, for any arbitrary relativistic observer,
only those stellar bodies are worthy to study which remain in stable position after
the application of fluctuations. Thus, the stability problem is among the burning issues for
the relativistic astrophysicists. With this intentions, we analyze
the stable regimes of our under considerations stellar models by applying the technique developed by
\cite{zs49}. This scheme could be useful to study the phenomenon of cracking. According to this scheme,
the stellar system must
have the ranges of $v^2_{sr}$ and $v^2_{st}$ between $0$ and $1$, e.g., in
the closed interval $[0,1]$, here $v_{sr}$ and $v_{st}$ represent the radial
and transverse sound speeds, respectively. These variables are defined as follows
$$\frac{{d{P_r}}}{{d\rho }} = v_{sr}^2,\quad \frac{{d{P_t}}}{{d\rho }} = v_{st}^2.$$
Now, using Eqs.(\ref{ro})-(\ref{pt}), their values are calculated as
\begin{align}\nonumber
v_{sr}=\sqrt{\frac{\chi}{\Psi}},\quad v_{st}=\sqrt{\frac{\phi}{\Psi}},
\end{align}
where
\begin{align}\nonumber
\chi&=r\{4r^3(B^2(1+\epsilon)+A^2(1+2\epsilon+Br^2(1+\epsilon))-AB(3+2\epsilon+Br^2(1+\epsilon)))f'+r\{2(1\\\nonumber
&+\epsilon)+Br^2(2+\epsilon)
-3Ar^2(2+3\epsilon)\}R'^2fi'''+R'\{(6Br^2-2+2B^2r^4-2\epsilon+3Br^2\epsilon+2B^2\\\nonumber
&\times r^4\epsilon+2A^2r^4(2+3\epsilon)-Ar^2
(8+11\epsilon+2Br^2(3+2\epsilon))-\exp(Ar^2)r^2(1+\epsilon)R\}fi''+3r^2\\\nonumber
&\times (2+3\epsilon)R''fi'''\}+rfi''\{(2(1+\epsilon)+Br^2(2+\epsilon)-3Ar^2(2+3\epsilon))R''+r(2+3\epsilon)R'''\}\\\label{x}
&+r^2(2+3\epsilon)R'^3fi''''\},\\\nonumber
\Psi&=-4[\exp(Ar^2)(1+\epsilon)-1-2\epsilon-B^2r^4\epsilon+A^2r^4\{2+(4-Br^2)\epsilon\}+Ar^2\{-1(4Br^2-2\\\nonumber
&+B^2r^4)\epsilon\}]fi'+r\{r(3(Br^2-2)\epsilon-4+3Ar^2(2+3\epsilon))R'^2fi'''+R'((2+14Ar^2-4A^2r^4+2\epsilon\\\nonumber
&+23Ar^2(1+\epsilon)
fi''=3r^2(2+3\epsilon)R''fi''')+rfi''((3Br^2\epsilon-4-6\epsilon+3Ar^2(2+3\epsilon))R''-r(2\\\label{y}
&+3\epsilon)R''')-r^2(2+3\epsilon)R'^2fi'''')\},\\\nonumber
\phi&=4[4\{1+2\epsilon+B^2r^4\epsilon+A^2Br^6\epsilon-\exp(Ar^2)(1+2\epsilon)+Ar^2(1+2Br^2+2\epsilon-B^2r^4\epsilon)\}fi'\\\nonumber
&+r\{-r(4+3(2+Ar^2)\epsilon)+Br^2(2+\epsilon)\}R'^2fi'''+R'\{(2+8Ar^2-6Br^2+4ABr^4+2\epsilon\\\nonumber
&+11Ar^2\epsilon-3Br^2\epsilon+2A^2r^4\epsilon+2B^2r^4\epsilon
+\exp(Ar^2)(2+4\epsilon)+\exp(Ar^2)r^2(1+\epsilon)R)fi''\\\label{z}
&+3r^2\epsilon R''fi'''\}-rfi''\{(4+6\epsilon+3Ar^2\epsilon+Br^2(2+\epsilon))R''-r\epsilon R'''\}+r^2\epsilon R'^3fi''''].
\end{align}
To reach the stable window, the transverse and radial speeds of the relativistic spherical
manifolds must obey $0 \le v_{sr}^2 \le 1$ and $0 \le v_{st}^2 \le 1$.
\begin{figure} \centering
\epsfig{file=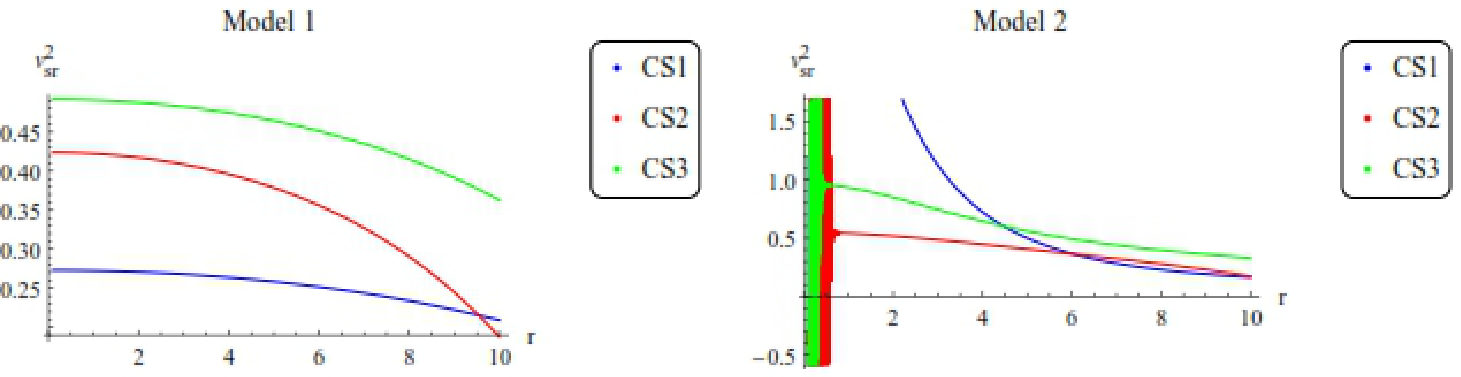,width=0.7\linewidth}
\caption{Change in $v_{sr}^2$ versus $r (km)$ of CS1, CS2 and CS3.}\label{vsr}
\epsfig{file=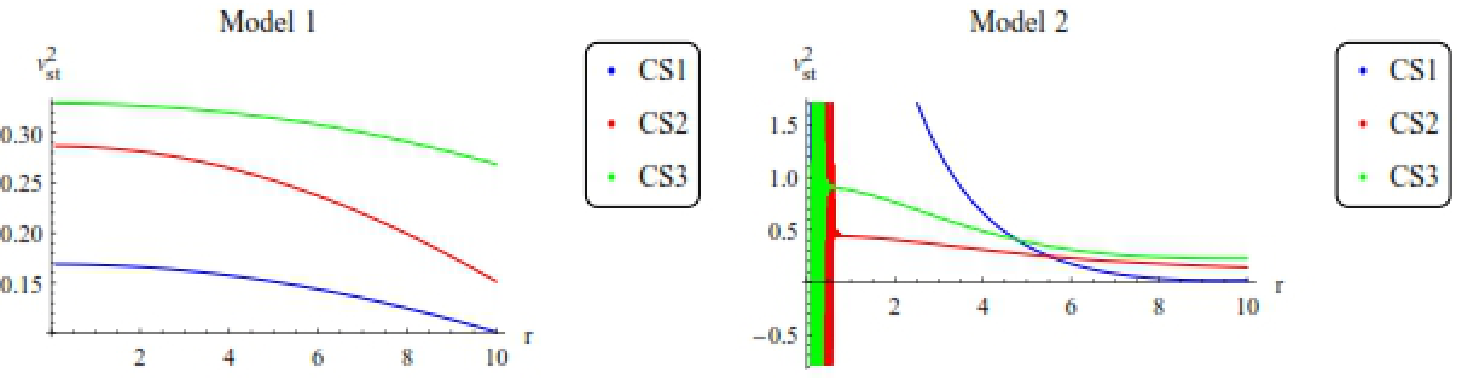,width=0.7\linewidth}
\caption{Change in $v_{st}^2$ versus $r (km)$ of CS1, CS2 and CS3.}\label{vst}
\end{figure}
We plotted some graphs by using Eqs.(\ref{model1})-(\ref{model3})
and (\ref{x})-(\ref{z}) along with Table \ref{table:1}.
One can clearly see from the Figures (\ref{vsr}) and (\ref{vst})
that $v_{sr}^2 $ and $v_{st}^2 $ are within the stability
range for all three observed compact stellar structures.
Furthermore, Figure (\ref{vstmvsr}) shows the stability
modes like
$$0< |v_{st}^2-v_{sr}^2|<1.$$
This indicate that our compact spherical geometries are within the stability
bounds even in the presence of extra degrees of freedom coming from (\ref{model1}) and (\ref{model3}) $f(R,T)$ models.
\begin{figure} \centering
\epsfig{file=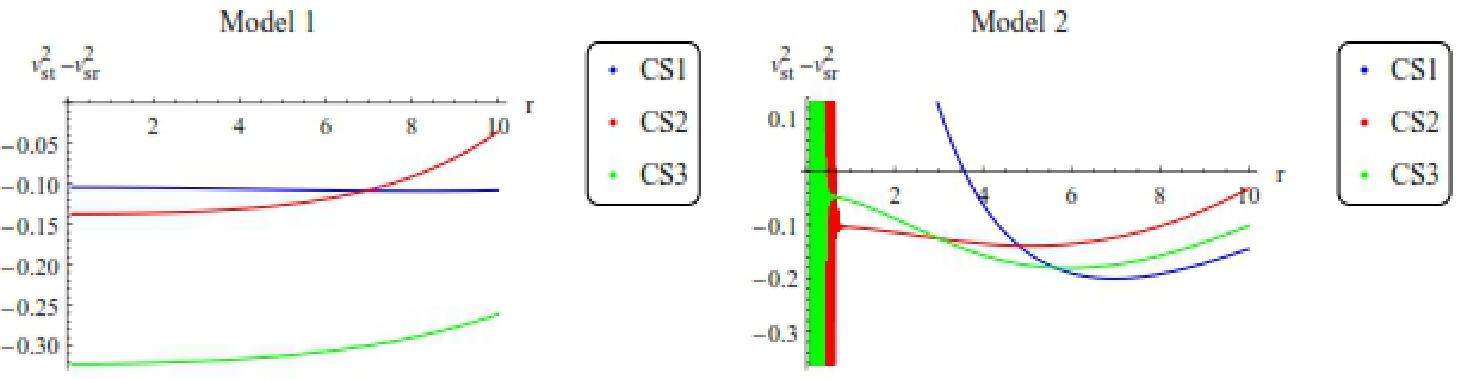,width=0.8\linewidth}
\caption{Change in $v_{st}^2 - v_{sr}^2$ versus $r (km)$ of CS1, CS2 and CS3.}\label{vstmvsr}
\epsfig{file=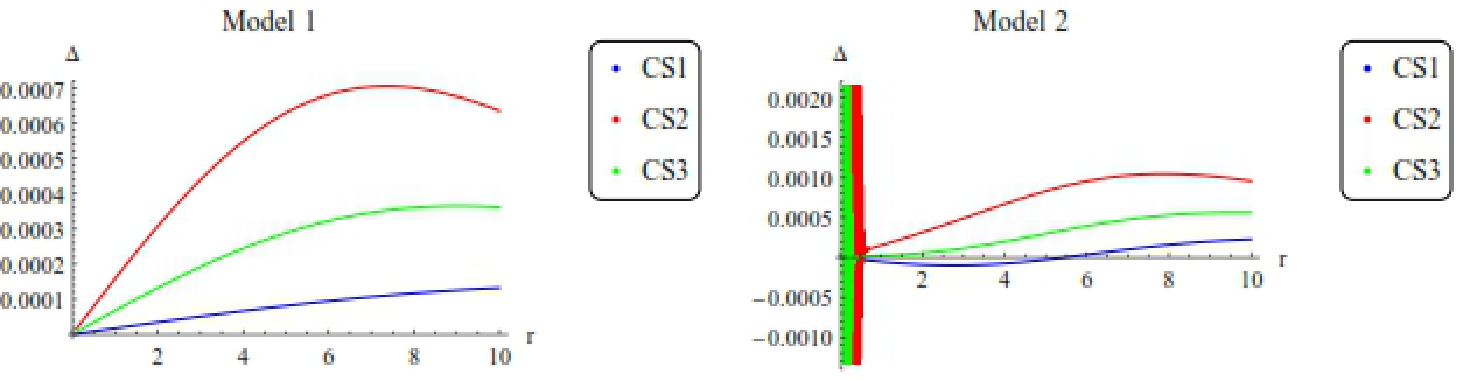,width=0.8\linewidth}
\caption{Change in $\Delta$ versus $r$.}\label{ani}
\end{figure}

\subsection{The Measurement of Anisotropy}

The extent of anisotropicity as well as its magnitude
within a local anisotropic fluid distributions can be defined as
\begin{align}\nonumber
\Delta  &\equiv \frac{2}{r}({P_t} - {P_r}).
\end{align}
Using Eqs.(\ref{ro})-(\ref{pt}), we can write $\Delta$ as
\begin{align}\nonumber
\Delta &=\frac{2\exp(-Ar^2)}{r^3(1+\epsilon)}[\{\exp(Ar^2)-(1+Ar^2-Br^2)(1+Br^2)\}fi'
+r\{-(1+Ar^2)\\\nonumber
&\times R'fi''+rfi''R''+rR'^2fi'''\}].
\end{align}
The above equation contains $fi$ terms. We use the two different viable $fi$ models
from Eqs.(\ref{model1})-(\ref{model3}) and get the
three different equations, each incorporating a particular $f(R,T)$ model corrections.
After using observational results of the compact stars given in Table \ref{table:1}, we terminated
with six set of equations. We check the behavior of anisotropy in
the degrees of anisotropicity in the background of relativistic compact objects, by plotting these equations. It can be seen from
the plot shown in Figure (\ref{ani}) that
the $\Delta$ is remained greater than zero for all the three relativistic compact
stellar structures which resulting that the influence of radial pressure, $P_r$
is lesser than that of transverse pressure, $P_t$. This outward directed measure of anisotropy
for these two viable models.

\section{Summary}

As the universe is accelerating, this ground reality attracts
researchers towards the extension of GR and
modified theories of gravity. The first modification to GR gives birth to $f(R)$ gravity,
which provides a good attention while the extension of this modification gives rise to $f(R,T)$ gravity and attain much attentions because of
some quantum effects arising in the theory. These modifications
appear in the field of low energy action for effective
quantum gravity theory.

In this work, we study the anisotropic relativistic compact
stellar objects with static and spherical structures. For this
purpose, we consider the relativistic compact stellar objects
whose interior geometry is based on anisotropy in $f(R,T)$ gravity. Krori and Barua \cite{zs43} proposed the
relativistic interiors of stellar bodies through specific metric combnations. With these
techniques, we connect this interior geometry with the exterior
Schwarzschild geometry and find the constant (in term of mass and
radius) of interior metric over the boundary. After this, we used
some observational data from which we find the numerical values
of these constants. Then we put these numerical values in our
calculations and plots our results.

In this paper, we have used BK solution, according to which $a(r) =B r^2 +C$ and $b(r)=A r^2$, where
$A,~B$ and $C$ are the three constant numbers which can be evaluated depending on several physical
requirements. Such solutions \cite{zs43} are claimed to be singularity-free
for the static spherical systems in the background of GR. Furthermore,
this solution is asserted to be regular everywhere and the matter variables, like
mass density, pressure etc are finite all over within the relativistic spherical system.

It is of our interest to check the outcomes and the predictions of one
of the extended gravities, i.e., $f(R,T)$ theory regarding the existence and stability of spherical stars.
Therefore, we have explored the modified version of the TOV equation in the realm of couple of $f(R,T)$ models.
Figure 9 states that for the particular choices of $f(R,T)$ model parameters, there exists some eras under which system could attain
equilibrium condition by keeping all the forces sum to be zero. It is worthy to mention that
usual GR forces are being modified due to $f(R,T)$ models, thus producing some extra effects in the forces $F_g,~F_h$ and $F_a$.
The stability of our compact stars not only depends upon KB-solution but also on the choices of parameters involved in $f(R,T)$ models mentioned
in Eqs.(4.2) and (4.3). The difference of the squares of sound speeds, i.e, $v_{st}^2-v_{sr}^2$ has been found to be
within $[0,1]$, thus describing our relativistic stars to be in stable window with certain
values of parameters involved in the corresponding $f(R,T)$ models.

The energy density remains positive and maximum at the core of
compact stars. Energy conditions holds for all these three
compact stars and the radial as well as the transverse equation of state
parameter are in a usual range i.g.  $0 < \omega_i < 1$,
here $i = r, t$. This indicate that the interior structures
of these relativistic compacts stellar objects are composed
of normal ordinary matter. For all case, we find the anisotropy
directed outward, e.g., $p_t > p_r$ or $ \Delta> 0$.
Similarly, the transverse as well as the radial sound speed
for all these compact stars are in stable limits that
implies the stability of these spherical anisotropic
compact bodies in the realm of $f(R,T)$ gravity.

\end{document}